\documentclass[twocolumn,superscriptaddress,preprintnumbers,amsmath,amssymb,prx]{revtex4-1}

\newcommand{\IMSS}{Muon Science Laboratory and Condensed Matter Research Center, Institute of Materials Structure Science, High Energy Accelerator Research Organization (KEK-IMSS), Tsukuba, Ibaraki 305-0801, Japan}
\newcommand{\Sokendai}{Department of Materials Structure Science, The Graduate University for Advanced Studies (Sokendai), Tsukuba, Ibaraki 305-0801, Japan}
\newcommand{\PSI}{Laboratory for Muon Spin Spectroscopy, Paul Scherrer Institute, CH-5232 Villigen PSI, Switzerland}
\newcommand{\MSLTokyo}{Materials and Structures Laboratory, Tokyo Institute of Technology, Yokohama, Kanagawa 226-8503, Japan}
\newcommand{\MCES}{Materials Research Center for Element Strategy, Tokyo Institute of Technology (MCES), Yokohama, Kanagawa 226-8503, Japan}

\newcommand{\Magazine}[4]{#1 {\bf #2}, #3 (#4).}
\newcommand{\JACS}{J. Am. Chem. Soc.\xspace}
\newcommand{\PRL}{Phys.~Rev. Lett.\xspace}
\newcommand{\PRB}{Phys.~Rev. B\xspace}

\newcommand{\JPSJ}{J.~Phys. Soc. Jpn.\xspace}

\usepackage{txfonts}
\usepackage[dvips]{graphicx}
\usepackage{dcolumn} 
\usepackage{bm} 
\usepackage{xspace}
\usepackage{ulem} 
\usepackage{multirow}
\usepackage{tabularx}
\begin{document}

\author{M.~Hiraishi}
\affiliation{\IMSS}
\author{K.~M.~Kojima}
\thanks{Present address: Centre for Molecular and Materials Science, TRIUMF, Vancouver, BC V6T2A3, Canada}
\affiliation{\IMSS}\affiliation{\Sokendai}
\author{H.~Okabe}
\affiliation{\IMSS}\affiliation{\Sokendai}
\author{S.~Takeshita}
\affiliation{\IMSS}
\author{A.~Koda}
\affiliation{\IMSS}\affiliation{\Sokendai}
\author{R.~Kadono}
\affiliation{\IMSS}\affiliation{\Sokendai}
\author{R.~Khasanov}
\affiliation{\PSI}
\author{S.~Iimura}
\affiliation{\MSLTokyo}
\author{S.~Matsuishi}
\affiliation{\MCES}
\author{H.~Hosono}
\affiliation{\MSLTokyo}\affiliation{\MCES}

\title{Magnetism driven by strong electronic correlation in the heavily carrier-doped iron oxypnictide LaFeAsO$_{0.49}$H$_{0.51}$}
\begin{abstract}
The magnetism of the second antiferromagnetic phase (AF2) arising in the iron-based LaFeAsO$_{1-x}$H$_{x}$ superconductor for $x\gtrsim0.4$ was investigated by muon spin rotation measurements under hydrostatic pressure up to 2.6 GPa. The N\'eel temperature ($T_{\rm N}$) obtained for a sample with $x=0.51$ exhibits considerably greater sensitivity to pressure than that in the pristine antiferromagnetic phase (AF1, $x\lesssim0.06$).
Moreover, while the AF1 phase is always accompanied by the structural transition (from tetragonal to orthorhombic) at a temperature ($T_{\rm s}$) which is slightly higher than $T_{\rm N}$, the AF2 phase prevails at higher pressures above $\sim$1.5 GPa where the structural transition is suppressed ($T_{\rm s}=0$).
These features indicate that the microscopic origin of the AF2 phase is distinct from that of AF1, suggesting that electronic correlation plays important role in the former phase.
We argue that the orbital-selective Mott transition is a plausible scenario to account for the observed pressure dependence of $T_{\rm N}$ and $T_{\rm s}$ in the AF2 phase.

\end{abstract}

\maketitle

\section{Introduction}
Since the discovery of high-$T_c$ superconductivity in iron-based oxypnictides $Ln$FeAsO$_{1-x}$F$_x$ (where $Ln$ denotes a lanthanide)~\cite{Kamihara,Ren2008,Wang2008,Hosono_review}, the interplay between magnetism and superconductivity in iron-based compounds has been a fascinating topic. While iron is an essential element of the electrically conducting FeAs planes, it usually plays an antagonistic role against superconductivity by bringing about magnetism. As a matter of fact, these compounds in pristine conditions exhibit antiferromagnetic (AF) order below the N\'eel temperature ($T_{\rm N}$), where superconductivity emerges as the AF order is suppressed by carrier doping of the FeAs plane~\cite{Lumsden}.
\par
The emergence of high-$T_c$ superconductivity upon suppression of AF order bears remarkable similarity with that in cuprates, where the parent compounds are regarded as typical Mott insulators. Although the microscopic mechanism of high $T_c$ is still under debate, it seems now commonly presumed that the electronic correlation on the CuO$_2$ planes (i.e., the strong on-site Coulomb repulsion, which leads to the metal-insulator transition upon half filling of the Cu $e_g$ band) is the essential ingredient in cuprates~\cite{Keimer:15}. Meanwhile, the iron-based compounds are distinct from cuprates in that the pristine compounds exhibit {\sl metallic} AF order (or spin density wave). Moreover, the AF order is always accompanied by a structural transition at the temperature $T_{\rm s}$ which is slightly higher than $T_{\rm N}$, suggesting a correlation between magnetism and the orbital degrees of freedom. These observations lead to the suggestion that spin and/or orbital fluctuations enhanced by the specific Fermi surface topology mediate the Cooper pairing~\cite{Hosono_review}.

\begin{figure}[!b]
	\centering
	\includegraphics[width=0.65\linewidth,clip]{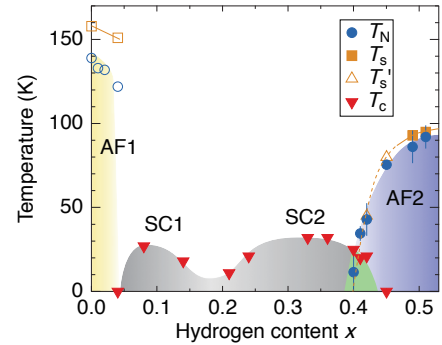}
	\caption{Electronic phase diagram of LaFeAsO$_{1-x}$H$_x$ along H content $x$, where AF1/2 and SC1/2 denote antiferromagnetic and superconducting phases~\cite{Hiraishi}. The experimentally determined N\'eel temperature $T_{\rm N}$, structural transition temperature $T_{\rm s}$, and superconducting transition temperature $T_{\rm c}$ are represented by circle, triangle, and inverted triangle symbols, respectively.}
	\label{Fig_PD}
\end{figure}

The recent development of carrier-doping technique using hydride ion ($Ln$FeAsO$_{1-x}$H$_x$) paved a path to large doping concentration up to $x\sim0.5$, providing opportunity to investigate the relationship between magnetism and superconductivity over unprecedented range of $x$~\cite{Hanna2011,Matsuishi_Ce,Iimura,LnFeAsOH_review}.
As shown in Fig.~\ref{Fig_PD}, the extended doping for the case of $Ln=$ La~\cite{Iimura} led to the discovery of a new superconducting phase (SC2) marked by the second peak of $T_{\rm c}$ around $x\sim0.36$ (with a dome-like $x$ dependence of $T_c$) and associated AF phase (AF2) that emerges for $x\ge0.4$ in place of the SC2 phase, establishing a novel bipartite phase diagram together with the pristine AF phase ($x\le0.05$, denoted AF1) and the known superconducting phase (SC1, accompanying another $T_c$ dome with a peak around $x\sim0.1$) which is separated by a valley of $T_{\rm c}$ near $x\sim0.2$~\cite{NMR_Fujiwara2013,Hiraishi,NMR_Sakurai2015}.

As inferred from the previous studies on a sample with $x=0.51$, the AF2 phase is characterized by a unique structural modulation and magnetic structure different from the AF1 phase~\cite{Hiraishi}.
Although the existence of the AF2 phase adjacent to SC2 in the bipartite phase diagram suggests a causal relationship between these two phases, the subtle difference between the AF1 and AF2 phases hints for possible distinction in the mechanism of superconductivity between the corresponding SC phases.
\par
According to the earlier resistivity measurements, the $T_{\rm c}$ domes of SC1 and SC2 phases tend to overlap when applying a hydrostatic pressure, merging into a single dome at 6~GPa, with a maximum $T_{\rm c}$ of 52~K ~\cite{Takahashi}.
Furthermore, it is inferred from the recent synchrotron X-ray diffraction measurements that the structural transition ($T_{\rm s} = 95$~K, for $x=0.51$) under ambient pressure is suppressed ($T_{\rm s}=0$) under a relatively low pressure of 1.5~GPa~\cite{Kobayashi}.
These features suggest a strong connection between $T_{\rm c}$ and the lattice structure, providing an important clue for understanding the mechanism behind high-$T_{\rm c}$ superconductivity in LaFeAsO$_{1-x}$H$_x$.
This connection naturally raises questions regarding the interrelationship between lattice structure and magnetism in the AF2 phase.
To address this issue, we conducted muon spin rotation ($\mu$SR) measurements under hydrostatic pressures on a LaFeAsO$_{1-x}$H$_x$ sample with $x=0.51$ situated in the AF2 phase.

\section{Experimental Methods}
A polycrystalline sample identical to that used for the X-ray diffraction experiment ($x=0.51$) \cite{Kobayashi} was adopted for the $\mu$SR experiment to avoid ambiguity coming from possible fluctuation of doping concentration, where the details of the sample preparation are reported in Ref.~\cite{Hanna2011}.
Conventional $\mu$SR measurements under a hydrostatic pressure were performed using the general purpose decay-channel (GPD) spectrometer of the Swiss-Muon-Source facility at the Paul Scherrer Institute, Switzerland.
A powder sample ($\sim$1.5~g) was pressurized within a cylindrical space with diameter of 5.9~mm using Daphne oil 7373 as pressure-transmitting medium.
The sample was sealed by a double-wall pressure-cell (PC) made of CuBe and MP35N (Ni, Co, Cr, and Mo alloy)~\cite{Khasanov}.
The exact pressure inside the cell was determined from the superconducting transition temperature $T_{\rm c}$ of a small piece of indium, which was also mounted on the same sample mount space~\cite{In_Tc}.
A muon beam with momentum of 99.25~MeV/$c$ was irradiated to penetrate the thick wall of the pressure-cell and to maximize the number of muons stopped in the sample space.
The pressure-cell was loaded onto a cryostat under He gas flow to monitor the time-dependent $\mu$SR spectra [positron decay asymmetry $A_z(t)$] under a zero (ZF) or transverse (TF, 5~mT) external field in the 5--140~K temperature range.
\begin{figure}[!t]
	\centering
	\includegraphics[width=\linewidth,clip]{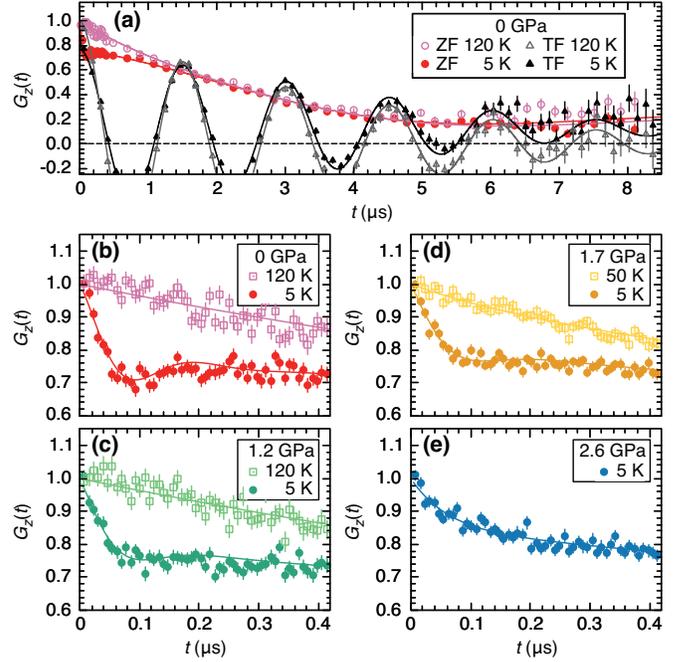}
	\caption{(a) Typical ZF- and TF-$\mu$SR spectra of LaFeAsO$_{1-x}$H$_x$ with $x=0.51$ (normalized to the value at $t=0$) measured at temperatures above (120~K) and below (5~K) $T_{\rm N}$ under ambient pressure. Spectra for TF=5~mT are partially represented for clarity. (b)--(e) ZF-$\mu$SR spectra measured under (b) ambient pressure, (c)~1.2~GPa, (d)~1.7~GPa, and (e)~2.6~GPa. Open and filled symbols represent the spectra above $T_{\rm N}$ and those at the lowest temperature, respectively. Solid curves are the best fits using Eq.~(\ref{eq1}).}
	\label{Fig1}
\end{figure}
\par

\section{Results}
In the high-pressure setup, the $\mu$SR spectra consist of two signal components, one corresponding to muons stopped in the sample and the other from the pressure-cell. In ZF, these are described by
\begin{eqnarray}
 A_0G_z(t)&=&A_{\rm smp}G^z_{\rm smp}(t)+A_{\rm pc}G^z_{\rm pc}(t),  \label{eq1}\\
G^z_{\rm pc}(t)&=&G^z_{\rm KT}(\Delta_{\rm ZF},t)\exp\left(-\lambda_{\rm pc} t\right),\nonumber
\end{eqnarray}
where $G^z_{\rm smp}(t)$ and $G^z_{\rm pc}(t)$ represent the time evolution of the muon spin polarization in the sample and in the pressure-cell, respectively, with their respective partial asymmetry being $A_{\rm smp}$ and $A_{\rm pc}$ ($A_0=A_{\rm smp} + A_{\rm pc}$).
$G^z_{\rm pc}(t)$ is known to be described by the static Kubo-Toyabe relaxation function $G^z_{\rm KT}(\Delta_{\rm ZF},t)$ multiplied by an exponential damping of rate $\lambda_{\rm pc}$ to empirically account for the depolarization in the cell, with the term $\Delta_{\rm ZF}$ representing the linewidth caused by nuclear magnetic moments~\cite{Khasanov}.
\par
Figure~\ref{Fig1}(a) shows typical examples of ZF- and TF-$\mu$SR spectra under ambient pressure.
The ZF spectrum at 5~K ($<T_{\rm N}$) is dominated by a slow Gaussian-like depolarization for $t\geq0.2$~$\mu$s, indicating that the lineshape at later times is predominantly determined by $G^z_{\rm KT}(\Delta_{\rm ZF},t)$ for muons stopped in the non-magnetic pressure cell. Meanwhile, the slow exponential-like depolarization at 120~K ($>T_{\rm N}$) represents the behavior of $G^z_{\rm smp}(t)$ overlapped with $G^z_{\rm KT}(\Delta_{\rm ZF},t)$, which originates from residual iron impurities known to exist in the present specimen~\cite{Iimura_private}.
Considering that the situation resembles the canonical dilute spin glass (e.g., $Au$Fe) \cite{Uemura:85}, we attribute this behavior to a spin glass-like impurity phase that coexists with the AF phase below $T_{\rm N}$.
\par
It is clear in magnified Figs.~\ref{Fig1}(b)--(e) that the spectra at 5~K exhibits fast damping precession under ambient pressure, and that the precession becomes obscure with increasing pressure, suggesting the decrease of internal field (and/or the fractional yield) of the AF phase probed by muon because of progressive suppression of magnetic correlation.
Considering these observations, the spectra were analyzed by $\chi$-square-minimization curve fitting using Eq.~(\ref{eq1}) and the following expression for $G^z_{\rm smp}(t)$:
\begin{align}
  G^z_{\rm smp}(t)=wG_{\rm mag}(t)+(1-w)G_{\rm sg}(t).\nonumber
\end{align}
Here, $G_{\rm mag}(t)$ represents the component exhibiting the relatively well-defined AF order with the volume fraction $w$, and $G_{\rm sg}(t)$ accounts for the remaining non-magnetic ($T_{\rm N}\sim0$) fraction dominated by spin glass-like behavior. Assuming quasi-static magnetism, $G_{\rm mag}(t)$ and $G_{\rm sg}(t)$ are approximated by
\begin{align}
  G_{\rm mag}(t)&=\frac{1}{3}+\frac{2}{3}\cos\left(2\pi ft+\phi\right)e^{-\Lambda t},\label{mag}\\
  G_{\rm sg}(t)&=\frac{1}{3}+\frac{2}{3}(1-\lambda t)e^{-\lambda t},\label{sg}
  \end{align}
where the first term represents the fraction of muons subject to the component of the quasi-static local field $B_{\rm loc}$ parallel to the initial muon spin direction $\hat{z}$ and is 1/3 for powder samples, and the second term represents that for $B_{\rm loc}\perp\hat{z}$. 
In $G_{\rm mag}(t)$, the muon spin exhibits precession with a frequency $f=\gamma_\mu B_{\rm loc}/2\pi$ (with $\gamma_\mu=135.538\times2\pi$~MHz/T being the muon gyro magnetic ratio) for non-zero $B_{\rm loc}$.
We adopted the Lorentzian Kubo-Toyabe function for $G_{\rm sg}(t)$ to describe the spin glass-like behavior \cite{Uemura:85}, where $\lambda$ was $\sim0.5~\mu{\rm s^{-1}}$ for the relevant temperature range.
In the quasi-static magnetic phase, the local field probed by muons is mainly determined by a vector sum of the magnetic dipolar field of the Fe atoms,
\begin{equation}
  B_{\rm loc}=|\sum_i\hat{{\bm A}_i}{\bm \mu}_i|,
  \label{Eq2}
\end{equation}
where ${\bm \mu}_i$ is the magnetic moment of the $i$-th Fe located at distance ${\bm r}_i=(x_i, y_i, z_i)$ from the muon site, and
\begin{equation*}
\hat{{\bm A}_i}=A_i^{\alpha\beta}=\frac{1}{r_i^3}\left(\frac{3\alpha_i\beta_i}{r^2_i}-\delta_{\alpha\beta}\right),\: (\alpha, \beta = x,y,z)
\end{equation*}
is the dipolar tensor. Although $B_{\rm loc}$ is a scalar quantity, it provides a strong criterion to verify the consistency among muon site(s), magnetic structure, and the Fe-moment magnitude inferred from other experimental techniques (see below). In the following curve-fit analysis, the fraction of muons stopped in the sample, $f_{\rm s}\equiv A_{\rm smp}/(A_{\rm smp}+A_{\rm pc})$, and the total asymmetry $A_0$ at the four applied pressures are fitted simultaneously in order to impose a common value for all temperatures, yielding $f_{\rm s}=0.45$--0.52 and $A_0\simeq0.28$.
For the signal coming from the pressure-cell, $\Delta_{\rm ZF}$ was fixed to the value obtained by interpolation at each temperature, and $\lambda_{\rm pc}$ was fixed to $0.04~\mu{\rm s}^{-1}$ since it is known to be almost unchanged down to 1~K~\cite{Khasanov} that is far below the lowest temperature attained in our study.
\par
\begin{figure}[!t]
	\centering
  \includegraphics[width=\linewidth,clip]{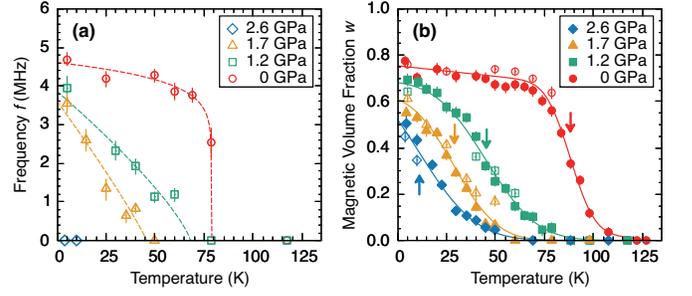}
	\caption{Temperature dependence of (a) $f$, and (b) $w$ under different pressures. Closed (open) symbols indicate values obtained from TF- (ZF-) $\mu$SR measurements, respectively. The dashed lines in (a) are a guide to the eye. The solid curves in (b) are the best-fit curves used to extract the magnetic transition temperature $T_{\rm N}$, denoted by solid arrows, as described in the text.}
	\label{Fig2}
\end{figure}

The temperature dependence of the frequency $f$ under four different pressures is shown in Fig.~\ref{Fig2}(a).
We note that $f$ was fixed to 0 in analyzing the spectrum under 2.6~GPa because no clear precession signal was discernible [see Fig.~\ref{Fig1}(e)].
The decrease of $f$ at the lowest-temperature with increasing applied pressure indicates that $B_{\rm loc}$ is reduced accordingly.
Assuming that the magnetic structure is unchanged, this suggests that the magnitude of the Fe moments [which is proportional to $B_{\rm loc}$, see Eq.~(\ref{Eq2})] decreases for increasing pressure (For the discussion on the possible broadening of $B_{\rm loc}$ induced by pressure, see Sect. \ref{Dis}).
\par
The fact that a long-lived precession with a frequency proportional to the external field ${\bm B}_0$ is observed above $T_{\rm N}$ for the entire asymmetry of TF-$\mu$SR spectra [including that corresponding to $A_{\rm smp}$, see Fig.~\ref{Fig1}(a)] indicates that the mean field in the spin glass-like phase is much weaker than ${\bm B}_0$, while those in the magnetic phase are depolarized rapidly because of the distribution of $B_{\rm loc}$ originated from the magnetic order,
\begin{align*}
  B_{\rm loc}=|\sum_i\hat{{\bm A}_i}{\bm \mu}_i+{\bm B}_0|.
\end{align*}
The upward shift of the 5~K spectrum seen in Fig.~\ref{Fig1}(a) derives from the first term in Eq.~(\ref{mag}), which reflects the muons subject to the total local field parallel to $\hat{z}$.
The $\mu$SR spectra under a TF of 5~mT were analyzed using Eq.~(\ref{eq1}) by replacing $G_{\rm pc}^z(t)$ and $G_{\rm sg}^z(t)$ respectively with
\begin{align*}
  G_{\rm pc}^z(t)&=\exp\left(-\lambda_{\rm pc} t\right)\exp\left(-\sigma_{\rm TF}^2t^2/2\right)\cos\left(\gamma_\mu B_0t+\phi\right),\\
  G_{\rm sg}^z(t)&=\exp\left(-\lambda t\right)\cos\left(\gamma_\mu B_0t+\phi\right),
\end{align*}
where $\sigma_{\rm TF}$ is the relaxation rate in TF-$\mu$SR measurements caused by the nuclear magnetic moments in the pressure-cell and $\phi$ is the initial phase of the precession.
$\sigma_{\rm TF}$ and $\lambda_{\rm pc}$ were also fixed to the values reported in Ref.~\cite{Khasanov}, as described above.
The temperature dependence of $w$ under different pressures is shown in Fig.~\ref{Fig2}(b), where the onset temperature decreases with increasing pressure.
At the lowest temperature, $w$ decreases upon increasing pressure, indicating that the volume fraction of the AF2 phase decreases. These results were used to determine the mean value and uncertainty of $T_{\rm N}$ (see below).

\section{Discussion}
\subsection{Muon Site}
In evaluating the magnetism of the AF2 phase based on the $\mu$SR results, it is important to have a good estimate of the muon site.
Since muons behave as a pseudo-hydrogen in matter, the variation in total energy upon inclusion of H estimated by density functional theory (DFT) calculations serves as a guide to narrow down the candidate muon sites.
The most probable site is inferred from the consistency of the $B_{\rm loc}$ calculated using Eq.~(\ref{Eq2}) for the candidate sites with that measured.
We calculated the total energy for interstitial H in the $Aem2$ orthorhombic phase using the OpenMX code, which is based on the generalized gradient approximation to DFT (DFT-GGA) and the norm-conserving pseudopotential method~\cite{openmx}.
We used a cutoff energy of 150~Ry and a $3\times3\times3$ mesh at the $K$-point with the experimentally obtained lattice constant~\cite{Hiraishi}.
\begin{figure}[t]
	\centering
	\includegraphics[width=\linewidth,clip]{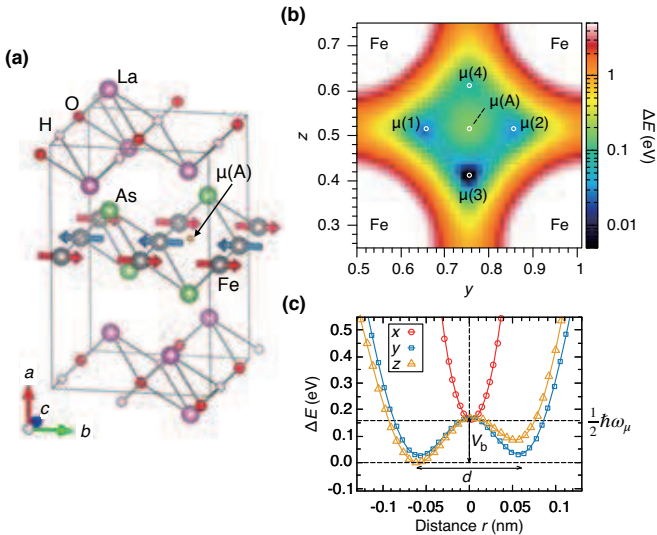}
	\caption{(a) Crystal and magnetic structure of LaFeAsO$_{0.49}$H$_{0.51}$. Possible muon site $\mu({\rm A})$ is represented by orange sphere. Red and blue arrows at Fe atoms represent magnetic moments along the $y$-axis. (b) Contour plot of $\Delta E$ in the $yz$-plane for $x=0.56$ in the $Aem2$ orthorhombic phase of LaFeAsO$_{1-x}$H$_x$ for $x=0.51$. Regions of $\Delta E>5$~eV are colored in white for clarity. $\mu(1)$--(4) and $\mu({\rm A})$ denote four $\Delta E$ minima and the central position (0.56, 0.75, 0.51). (c) $\Delta E$ profiles along the $x$-,$y$-, and $z$-axis passing through the $\mu({\rm A})$ site as functions of the distance $r$ from the $\mu({\rm A})$ site.}
	\label{Fig_Etot}
\end{figure}
\par
Figure~\ref{Fig_Etot}(b) shows the variation of the total energy $\Delta E=E_{\rm tot}({\bm r})-E_{\rm min}$ against the H position ${\bm r}$, where $E_{\rm min}$ is the global minimum of $E_{\rm tot}({\bm r})$.
As marked by $\mu(1)$--(4), four minima of $\Delta E$ are revealed around the central position $\mu({\rm A})$, which corresponds to the saddle point.
The slightly different values of $\Delta E$ at the minima may be attributed to the off-center deformation of the FeAs$_4$ tetrahedron in the $Aem2$ structure.
The distance from $\mu({\rm A})$ is 0.056~nm, 0.059~nm, and 0.053~nm for $\mu(1,2)$, $\mu(3)$, and $\mu(4)$, respectively.
Figure~\ref{Fig_Etot}(c) represents the $\Delta E$ profiles along the $x$-, $y$-, and $z$-axis passing through the $\mu({\rm A})$ site.
$\Delta E$ rapidly increases along the $x$-axis ($\perp$ to the FeAs plane) when increasing the distance $|r|$ from $\mu({\rm A})$, indicating that muons may be confined within the $yz$-plane.
However, along the $y$- and $z$-axis, $\Delta E$ exhibits an asymmetric double-well potential structure with maximum potential barrier $V_{\rm b}=169.1$~meV located between $\mu(3)$ and $\mu(4)$ at a distance $d=0.112$~nm.
In the harmonic approximation of the sinusoidal potential curve, the energy level splitting $\hbar\omega_\mu$ for the bound state muon is derived from the relation $\omega_\mu^2=2\pi^2V_{\rm b}/m_\mu d^2$, where $m_\mu=105.658$~MeV/$c^2$ is the muon mass.
The corresponding zero-point energy $\frac{1}{2}\hbar\omega_\mu$ in this approximation is estimated to be 156.5~meV, comparable to $V_{\rm b}$.
This suggests that the muon is virtually located at the $\mu({\rm A})$ site for a time longer than $ \omega_\mu^{-1}$, constituting a typical example of the isotope effect between muon and hydrogen.
\par
The muon site was identified by comparing the $B_{\rm loc}$ obtained from the ZF-$\mu$SR measurements in the magnetically ordered phase with that calculated using Eq.~(\ref{Eq2}) for the candidate sites, summing the Fe moments located within 10~nm from the muon site.
The $B_{\rm loc}^{\rm sim}$ values calculated at each muon site for $\mu({\rm A})$ and the $\mu(1)$--(4) minima, using the reported crystal and magnetic structure for $x=0$~\cite{Qureshi} and $x=0.51$~\cite{Hiraishi}, are summarized in Table.~\ref{table}.
For $x=0$, although $B_{\rm loc}^{\rm sim}$ agrees with the value obtained experimentally at $\mu({\rm A})$, it is larger than $B_{\rm loc}$ by a factor of 1.5$\sim$1.8 at the $\mu(1)$--(4) sites.
Similarly, for $x=0.51$, $B_{\rm loc}^{\rm sim}$ at $\mu(1)$--(4) is much larger than $B_{\rm loc}$, by a factor of $\sim$10.
Simulated value at $\mu({\rm A})$ for $x=0.51$ represents the range of $B_{\rm loc}^{\rm sim}$ within $\Delta r=5.8$~pm from $\mu({\rm A})$ position, where $\Delta r$ corresponds to the resultant mesh size for real-space in our DFT calculation.
Although $B_{\rm loc}^{\rm sim}$ changes steeply from 0 to 49.7~mT even for such a small mesh size (see Fig.~\ref{Fig_fld}(b)), its simple average within $\Delta r\simeq8$ pm yields $\sim$38~mT in close agreement with the experimental value.
These results indicate that the muon occupies $\mu({\rm A})$ for $x=0$ and 0.51 because of its small mass, unlike hydrogen for $x=0.1$~\cite{Yamaura2019}.

\begin{table}[!t]
	\caption{Simulated internal magnetic field $B_{\rm loc}^{\rm sim}$ at each muon site induced by the magnetic moments of the Fe atoms. The value of $B_{\rm loc}^{\rm sim}$ for $x=0.51$ at $\mu({\rm A})$ represents the range of $B_{\rm loc}^{\rm sim}$ around $\mu({\rm A})$ position within $\Delta r=5.8$~pm in $bc$-plane (see Fig.~\ref{Fig_fld}(b)).}
	\centering
	\vspace{1mm} {
    \renewcommand\arraystretch{1.3}
    \tabcolsep=2mm
    \begin{tabular}{c|cc}
		  \hline\hline
      \multirow{3}{*}{Muon site} & \multicolumn{2}{c}{Magnetic field (mT)}\\
      \vspace{-1.2mm}
		  & $x=0$ & $x=0.51$\\
      & ($|{\bm m}|=0.63\mu_{\rm B}$~\cite{Qureshi}) & ($|{\bm m}|=1.21\mu_{\rm B}$~\cite{Hiraishi})\\\hline
      $\mu({\rm A})$ & 144.3 & 0--49.7\\
			$\mu$(1), $\mu$(2) & 259.2 & 467.0\\
      $\mu$(3) & 306.6 & 371.9\\
			$\mu$(4) & 306.6 & 319.9\\
      \hline
      Experiment & 173.2~\cite{Luetkens2009,MaeterPRB2009} & 37.7$\pm$0.5~\cite{Hiraishi}\\\hline\hline
 	  \end{tabular}
  }
	\label{table}
\end{table}

\subsection{Magnitude and Distribution of ${\bm B}_{\rm loc}$ below ${\bm T_{\rm N}}$}\label{Dis}

\begin{figure}[!t]
	\centering
	\includegraphics[width=\linewidth,clip]{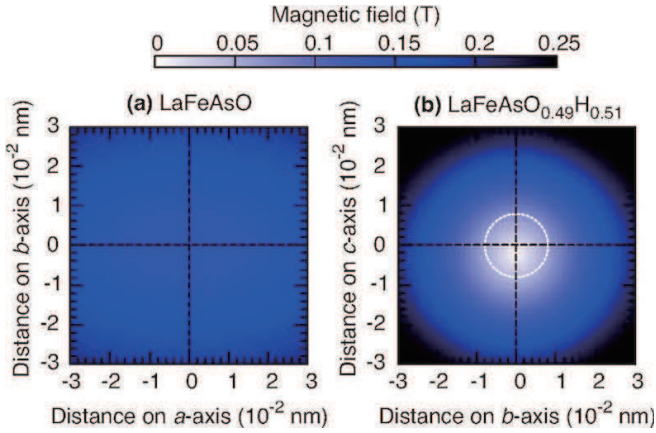}
	\caption{(a) In-plane $B_{\rm loc}^{\rm sim}$ magnetic field distribution around the muon site within $0.03\times0.03$~nm at (a) $z=0.573$ for LaFeAsO, and (b) $x=0.56$ for LaFeAsO$_{0.49}$H$_{0.51}$. Regions where the magnetic field is $> 0.25$~T are colored in black for clarity. The white dotted circle in (b) represents the region within
$\Delta r\simeq 8$ pm where the field is averaged to yield $B_{\rm loc}^{\rm sim}\simeq38$ mT. }
	\label{Fig_fld}
\end{figure}

Figure~\ref{Fig_fld}(a) and (b) represent the simulated internal magnetic field distribution around muon site $\mu({\rm A})$ in the plane parallel to the FeAs layer for $x=0$ and $x=0.51$, respectively.
Although the profile of the field around muon site $\mu({\rm A})$ (at the center of the graph) is nearly independent of position for $x=0$, a steep profile is revealed for $x=0.51$.
This indicates that a tiny displacement of the muon site does not affect the field profile probed by muons for $x=0$, whereas a strong dependence is expected for $x=0.51$.
This is especially important when considering that a muon site displacement is more probable for $x=0.51$ than for $x=0$ because the substituted hydrogen randomly occupies the oxygen site.
Thus, for $x=0.51$, muons may probe the broad field profile, causing the fast depolarization spectrum below $T_{\rm N}$ seen in Fig.~\ref{Fig1}(b).
The underestimated value of $B_{\rm loc}^{\rm sim}$ at $\mu({\rm A})$ for $x=0.51$ may be attributed to this broad profile.
\par
We examined the influence of pressure to the local field by calculating $B_{\rm loc}^{\rm sim}$ for the lattice constants reduced by external pressure reported in Ref.~\cite{Kobayashi}. According to the recent NMR experiment, the AF2 phase is stable below 2~GPa because of a large gap~\cite{Takeuchi_PRB2019}, suggesting that the magnetic structure is unchanged. Under the further assumption that the muon site is also intact with pressure, $B_{\rm loc}^{\rm sim}$ at 2~GPa increases by $\sim$2~\%, as expected from the definition of $B_{\rm loc}^{\rm sim}$ provided by Eq.~(\ref{Eq2}).
This trend is opposite to the experimental results, as shown in Fig.~\ref{Fig2}(a), necessitating other causes for the observed decrease of $B_{\rm loc}$ with increasing pressure. In this regard, it is interesting to note that a reduction of the Fe-moment magnitude under pressure is predicted by theoretical studies of LaFeAsO~\cite{Opahle,Labegue,Yang}, which is understood as a result of increased energy band width. Our estimation indicates that the reduction of the Fe moment by $\sim$25~\% is sufficient to account for the experimental result around 2~GPa.

We also draw the attention to that the $\mu$SR spectrum for $x=0.51$ at the lowest temperature and under ambient pressure is dominated by one precession term with fast depolarization. This is in sharp contrast to the $x=0$ case, where the spectrum at the lowest temperature is well reproduced by the sum of higher (lower) frequency term of $f\sim$23~MHz (3~MHz) with the fractional ratio of 7:3~\cite{Luetkens2009,MaeterPRB2009,Klauss2008,Renzi}.
This indicates the presence of two magnetically inequivalent muon sites. The second lowest energy site, $\mu({\rm B})=(0.09, 0.37, 0.11)$ for $x=0.51$, is assumed to be in the La-O/H layer with distance $r_{\rm O\mu}\simeq0.13$~nm from the nearest oxygen. This suggests the formation of a local state bound to oxygen, as that empirically established in many oxides (typical $r_{\rm O\mu}\simeq0.1$~nm).

\par
We calculated the local field at the $\mu({\rm B})$ site, finding $B_{\rm loc}^{\rm sim}=0.8$~mT. The smaller $B_{\rm loc}$ is ascribed to the greater $|{\bm r}_i|$ for the $\mu({\rm B})$ site from Fe moments [see Eq.~(\ref{Eq2})], where the distance from the nearest Fe atom to the $\mu$(A) and $\mu$(B) sites are 0.20~nm and 0.37~nm, respectively. Although the corresponding signal at $f=0.1$~MHz has a magnitude that can be detected by conventional $\mu$SR measurements, in our experimental conditions ($f_{\rm s}\sim$0.5) the low fractional ratio (0.3--0.4) for the $\mu({\rm B})$ site~\cite{MaeterPRB2009,Klauss2008,Renzi} and a signal-to-noise ratio of $\sim$1 impede separating this component from the slowly depolarizing ones, i.e., $G_{\rm sg}(t)$ and/or $G_{\rm pc}^z(t)$ in Eq.~(\ref{eq1}).
This may result in underestimating $w$ in Fig.~\ref{Fig2}(b).
Actually, the nonmagnetic volume fraction deduced under ambient pressure at $T\rightarrow0=1-w(T=0)\simeq0.25$, could correspond to the fraction $f_{\rm s}(B)$ of muons stopped at $\mu$(B). Supporting this, we note that its value is comparable with the expected value of $f_{\rm s}(B)$ in the present setup, $f_{\rm s}(B)=(0.3$--$0.4)\times f_{\rm s}=0.13$--0.18 with $f_{\rm s}=0.45$.
We also refer to the possibility that $B_{\rm loc}$ exhibits broader distribution under pressure, which will result in the absence of any detectable oscillation for the spectrum stems from the $\mu$(B) site.

\subsection{Magnetism vs Lattice Structure}
\begin{figure}[!t]
	\centering
	\includegraphics[width=0.7\linewidth,clip]{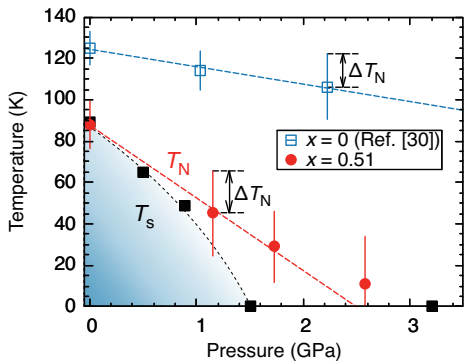}
	\caption{$T_{\rm N}$ , $\Delta T_{\rm N}$, as a function of external pressure, and $T_{\rm s}$ data from Ref.~\cite{Kobayashi}. The transition width $\Delta T_{\rm N}$ is represented as error bars for $T_{\rm N}$. Data represented by blue square symbols are quoted from Ref.~\cite{Renzi}. The dashed lines and the blue-colored region are guides to the eye.}
	\label{Fig5}
\end{figure}

The solid curves of Fig.~\ref{Fig2}(b) are the best fits using an equation $$w(T)=\frac{1}{2}w(0)\left[1-{\rm erf}\left(\frac{T-T_{\rm N}}{\sqrt{2}\Delta_{T_{\rm N}}}\right)\right],$$ in which a Gaussian distribution of width $\Delta_{T_{\rm N}}$ is assumed around the average transition temperature $T_{\rm N} $ (a linear term was added only for ambient pressure data to account for the gradual increase with decreasing temperature)~\cite{Sanna2009}.
The obtained $T_{\rm N}$ and that for $x=0$~\cite{Renzi} are shown in Fig.~\ref{Fig5} as a function of pressure, along with the structural transition temperature $T_{\rm s}$ determined for the same sample in Ref.~\cite{Kobayashi}.
The width $\Delta_{T_{\rm N}}$ is represented as error bars for $T_{\rm N}$, although $T_{\rm N}$ itself is well determined within an error of $\sim$1~K, except for the highest pressure.
The large error bars resulting for $T_{\rm N}(2.6$~GPa)$\sim$6~K may originate from the strong temperature-dependent behavior of $w$ in the lowest temperature region seen in Fig.~\ref{Fig2}(b).

It is remarkable that the structural transition to the orthorhombic phase for decreasing temperature is suppressed near 1.5~GPa ($T_{\rm s}\rightarrow0)$, whereas the AF2 phase survives even under 2.6~GPa.
This is in sharp contrast with the AF1 phase, in which $T_{\rm N}$ is always below $T_{\rm s}$, indicating that the magnetic order of the AF2 is induced by a purely electronic mechanism.
According to a theoretical study based on molecular orbitals~\cite{Iimura2}, the electronic state of the AF2 phase is understood through an orbital-selective Mott transition, where Fe-$3d_{xy}$ becomes half-filled when increasing $x$.
This situation is similar to that of the $e_g$ orbital in pristine cuprate compounds, implying that the AF order of the AF2 phase is induced by electronic correlation.
The fact that $T_{\rm N}$ is independent of $T_{\rm s}$ in the AF2 phase supports above expectation.
The pressure dependence of $T_{\rm N}$ is also understood within this scenario (see below).
\par
Here, it may be worth mentioning that the appearance of the AF order that precedes structural transition with reducing temperature bears a remarkable similarity with the so called electronic nematicity revealed in the BaFe$_2$As$_2$ (122) family compounds, where the isovalent substitution of As with P induces a unidirectional self-organized state that breaks the rotational symmetry of the underlying lattice above $T_{\rm s}$~\cite{Nematic_2012}. In addition, the coexistence of AF2 phase and SC2 phase observed over a finite doping range of $0.4\le x \le0.45$~\cite{Hiraishi} comprises yet another parallelism with the 122 family~\cite{Ba122}, hinting for the importance of electronic correlation in the latter compounds.

\par

The sensitivity of magnetism to pressure in the AF2 phase is quantitatively described by the gradient ${\rm d}T_{\rm N}/{\rm d}p=-35.5\pm0.6$~K$\cdot$GPa$^{-1}$, which is much greater than the $-8.5\pm0.1$~K$\cdot$GPa$^{-1}$ of the AF1 phase ($0\le x\le0.06$)~\cite{Renzi}, indicating that the AF2 phase is more susceptible to pressure than AF1.
This contrast may originate from the different mechanism of magnetic order.
In the AF1 phase, the AF order is induced by nesting of the Fermi surface, as is concluded by a theoretical study on LaFeAsO ($x=0$, AF1 phase) reporting that the nesting condition is almost unchanged between 0.1419~nm$^3$ and 0.120~nm$^3$ (corresponding to $-7$~GPa and 10~GPa), indicating the robustness of the AF1 phase against pressure~\cite{Opahle}.
On the other hand, monotonic degradation of the nesting with doping due to the expansion of the electron Fermi surface at the $M$ point~\cite{SuzukiPRL2014} disfavors the similar scenario for the AF2 phase.
\par
It is reported that the energy gap between bonding and antibonding orbitals upon Fe 3$d$-As 4$p$ hybridization in the low $x$-region decreases when the height of the As ion from Fe plane $h_{\rm As}$ increases.
As $h_{\rm As}$ increases (which is equivalent to an increase in $x$), the non-bonding Fe-3$d_{xy}$ orbital becomes half-filled, resulting in the orbital-selective Mott state.
Because external pressure induces a considerable decrease of $h_{\rm As}$ in LaFeAsO$_{1-x}$H$_x$~\cite{Kobayashi}, the fragility of the AF2 phase against pressure is readily understood within the scenario of orbital-selective Mott transition~\cite{Iimura2}.
In fact, for $x=0.51$ and at ambient pressure, $h_{\rm As}=0.1413$~nm, decreasing to 0.1375~nm under 2.2~GPa. This value is comparable to that for $x\sim0.4$, where the AF2 phase is nearly suppressed~\cite{Hiraishi,Kobayashi}.

Finally, we note that the strong pressure dependence of $T_{\rm N}$ combined with spatial inhomogeneity of pressure in the sample space originating from a partial non-hydrostaticity may also contribute to the non-magnetic phase ($1-w$) below $T_{\rm N}$, the confirmation of which remains as a future task using more refined $\mu$SR sample environment for high pressure.

\section{Summary and Conclusion}
To summarize, our $\mu$SR study of LaFeAsO$_{1-x}$H$_x$ with $x=0.51$ under external pressure revealed that the AF2 phase survives under a pressure as high as 2.6~GPa, far beyond the pressure where the structural transition to the orthorhombic phase is suppressed.
The AF2 phase for $x=0.51$ is more susceptible to pressure than the AF1 phase for $x=0$, suggesting a different magnetic ordering mechanism.
Considering theoretical works, the AF1 phase is robust against external pressure because the nesting of the Fermi surface that induces it is nearly independent of pressure. In contrast, the AF2 phase is understood through the orbital-selective Mott state, in which the height parameter $h_{\rm As}$ plays an essential role.
Because $h_{\rm As}$ decreases when applying external pressure, the AF2 phase is sensitive to the latter.

\section*{Acknowledgment}
This work was supported by the MEXT Elements Strategy Initiative to Form Core Research Centers (Grant Number JPMXP0112101001).
The $\mu$SR measurements were carried out at the Paul Scherrer Institute (PSI), Switzerland, under user program (Proposal No. 20152061).
We would like to thank the staff of the PSI for their technical support during experiments.


\end{document}